\documentclass{article}

\usepackage{arxiv}

\usepackage{fancyhdr}
\renewcommand{\today}{}

\usepackage[utf8]{inputenc} % allow utf-8 input
\usepackage[T1]{fontenc}    % use 8-bit T1 fonts
\usepackage{url}            % simple URL typesetting
\usepackage{booktabs}% professional-quality tables
\usepackage{multirow}
\usepackage{amsfonts}       % blackboard math symbols
\usepackage{nicefrac}       % compact symbols for 1/2, etc.
\usepackage{microtype}      % microtypography
\usepackage{lipsum}		% Can be removed after putting your text content
\usepackage{graphicx}
\usepackage{natbib}
\usepackage{doi}
\usepackage{graphicx}% Include figure files
\usepackage{amsmath}
\usepackage{dcolumn}% Align table columns on decimal point
\usepackage{bm}% bold math
\usepackage{natbib}
\usepackage{subcaption}
\usepackage{hyperref}% add hypertext capabilities
\hypersetup{
  colorlinks   = true, %Colours links instead of ugly boxes
  urlcolor     = blue, %Colour for external hyperlinks
  linkcolor    = blue, %Colour of internal links
  citecolor   = blue %Colour of citations
}
\usepackage[mathlines]{lineno}% Enable numbering of text and display math
\title{UNSCREENING OF $f(R)$ GRAVITY NEAR THE GALACTIC CENTRE BLACK HOLE: TESTABILITY THROUGH PERICENTRE SHIFT BELOW S0-2's ORBIT}

\author{ \href{https://orcid.org/0000-0003-4301-3496}{\includegraphics[scale=0.06]{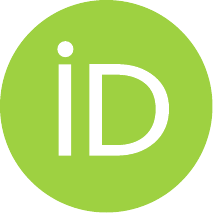}\hspace{1mm}Debojit Paul}\\
	Department of Physics\\
	Gauhati University\\
	Jalukbari, Guwahati-781014, Assam, India \\
	\texttt{debojit@gauhati.ac.in} \\
	\And
	\href{https://orcid.org/0000-0002-2880-4284}{\includegraphics[scale=0.06]{orcid.pdf}\hspace{1mm}Sanjeev Kalita} \\
	Department of Physics\\
	Gauhati University\\
	Jalukbari, Guwahati-781014, Assam, India \\
	\texttt{sanjeev@gauhati.ac.in} \\
    \And
    \href{https://orcid.org/0000-0002-5833-376X}{\includegraphics[scale=0.06]{orcid.pdf}\hspace{1mm}Abhijit Talukdar} \\
	Department of Physics\\
	Gauhati University\\
	Jalukbari, Guwahati-781014, Assam, India \\
	\texttt{abhijit@gauhati.ac.in} \\
}

\begin{document}
\maketitle
\begin{abstract}
General Relativity (GR) has been tested extensively in the solar system and is being tested in the new environment of the Galactic Centre (GC) black hole where the dimensionless gravitational potential ($GM/c^2r$) is 100 times stronger than the one encountered in solar system. Therefore, the neighbourhood of the GC black hole is a naive opportunity to test modified theories of gravity. In this work, effect of $f(R)$ gravity near the black hole is studied. The difference of pericentre shift between GR and $f(R)$ gravity is studied for compact orbits having semi-major axis equal to and below $a=1000$ au  (S0-2 like orbits). In a model dependent approach, we choose $f(R) \propto R^2$ (power law gravity) model which is cosmologically motivated and study the deviation in orbital pericentre shift for both zero spin and non-zero spin of the black hole. It is found that effect of $f(R)$ gravity becomes prominent for compact orbits. In model independent approach to $f(R)$ gravity with the generic scalaron fields ($\psi=f'(R)$), we extract the parameters of $f(R)$ gravity from the current bounds on Parametrised Post Newtonian (PPN) parameters ($\gamma , \beta$) near the GC black hole. The screening of $f(R)$ gravity is also investigated for these bounds on PPN parameters. It has been found that sufficiently massive scalarons ($10^{-16}$ eV) are completely screened but light and intermediate mass scalarons ($10^{-22}$ eV and $10^{-19}$ eV) are unscreened towards S0-2 like orbits as well as in the orbit of the newly discovered short period star S4716 ($a=407$ au). The possibility of detection of the $f(R)$ gravity effects due to these unscreened scalarons is forecasted with existing and upcoming astrometric capabilities of Extremely Large Telescopes (ELTs).
\end{abstract}

% keywords can be removed

\section{Introduction}
The supermassive black hole, Sgr A* of mass $M_{BH}=({4.261 \pm 0.012}) \times 10^6\ M_{\odot}$\citep{2020A&A...636L...5G} at the centre of our galaxy presents an exquisite laboratory for understanding gravity. General Relativity (GR) has been found to be a consistent theory of gravity through solar system, binary pulsar and gravitational wave tests\citep{2001LRR.....4....4W,1974ApJ...191L..59H,2016PhRvL.116f1102A}. However, some fundamental problems are yet to be addressed within the regime of the theory. They are: (a) Occurrence of black hole/cosmological singularity which naturally demands quantum corrections to the theory (b) existence of mysterious dark matter and dark energy, the former being essential for large scale structure formation and the latter being responsible for accelerated cosmic expansion. Within the framework of GR these are exotic matter particles or field/fluids respectively whose nature is not yet known.

\par Although a quantum theory of gravity is still under development many alternative theories of gravity are being exercised to explain cosmological observation without introducing dark matter and dark energy\citep{2021Univ....7..407B}. Many of these theories contain additional and massive gravitational degrees of freedom described by fields emerging from geometrical correction to gravitational field equations of GR\citep{Faraoni:2010pgm}. $f(R)$ gravity which emerges due to replacement of the Ricci scalar $R$ by a general function $f(R)$ in the Einstein-Hilbert action is among several mathematically consistent metric theories of gravity which are extensively discussed in the context of accelerated expansion\citep{2000PhRvL..85.2236B,2001PhRvD..63f3504E,2007PhRvD..76f4004H,2007JETPL..86..157S}. The Galactic Centre (GC) black hole, due to presence of compact stellar orbits and dark mass distributions\citep{2008ApJ...689.1044G,2009ApJ...707L.114G,2009ApJ...693L..35M,2017arXiv171106389D,2017PhRvL.118u1101H} is an excellent laboratory to test these alternative theories\citep{2021Univ....7..407B,2017PhRvL.118u1101H,2022MNRAS.513L...6Z,2019IJMPD..2841003Z,2012PhRvD..85l4004B,2021PhRvD.104j1502D,2020ApJ...893...31K,2021ApJ...909..189K}.   

\par In cosmology $f(R)$ gravity was first introduced by \cite{1980PhLB...91...99S} to explain primordial cosmic inflation via geometrical modification of general relativity which replaces standard scalar fields used to generate the inflation and density perturbations. This model invokes a correction to GR in the form of $f(R)=R^2$. A positive exponent of the Ricci scalar indicates that these corrections are used for high curvature modification to GR. After the discovery of late time cosmic acceleration \citep{1998AJ....116.1009R,1999ApJ...517..565P}, $f(R)$ gravity models of the type $f(R)=R+\alpha R^{-m}(m>0)$ and $f(R) \propto R+[(1+{R^2/R_0^2)}^{-n}-1](n,R_0>0)$ (Hu-Sawicki type) were used to explain the late time cosmic acceleration due to low curvature (large scale) modification to GR\citep{2007PhRvD..76f4004H,2007JETPL..86..157S,2008CQGra..25t5002S,2003astro.ph..3041C}. These modifications replaced exotic negative pressure sources including the cosmological constant, known as dark energy, as candidates for late time acceleration of the universe. It is now generally anticipated that both primordial and late time cosmic acceleration can be accommodated within modified theories of gravity. $f(R)$ gravity also modifies the matter sector of the universe which is otherwise dominated by mysterious non-baryonic dark matter. Missing matter in galaxies accounting for flat rotation curves, structure of self-gravitating systems such as stars, galaxies and their clusters and dynamics of collisionless systems have been successfully explained by $f(R)$ gravity without incorporating exotic matter particles\citep{Capozziello_2017,2012arXiv1202.0394D,2012PhRvD..85d4022C}. But, in most of the $f(R)$ models, it is not clear whether they satisfy the weak solar system tests and large-scale late time expansion of the universe simultaneously. This happens because $f(R)$ models introduce scalar degrees of freedom or scalarons that manifest as a long range fifth force. The scalaron degree of freedom is described by the scalar field $\psi=f'(R)=df(R)/dR$. A cannonically normalised scalar field is defined as $\sqrt{\frac{3}{2\kappa^2}}\ln |f’(R)|$ where, $\kappa^2=8\pi G$. It modifies the weak field metric and hence affects the solar system tests \citep{2010deot.book.....R}. In order to resolve this issue, \cite{2004PhRvD..69d4026K} introduced an approach known as chameleon mechanism which gives rise to screening of modified gravity \citep{2012A&G....53d..37L}. This mechanism gives dynamic characteristics to the scalarons. The scalarons in high density region become massive due to which their effect is completely screened in small scales. On the other hand, in low density regions these scalarons become light resulting in their leakage to larger scales. \cite{2020ApJ...893...31K} presented a stringent bound on mass of the scalarons near GC black hole as $M_{\psi}=(10^{-22}-10^{-16})$ eV. In this work the masses of the scalarons which are unscreened or screened near the GC black hole are extracted to understand the scale where gravity is likely modified in the black hole environment.

\par Pericentre shift of compact stellar orbits near the GC black hole is a direct astronomical probe for testing any gravitational theory\citep{2020ApJ...893...31K,2021ApJ...909..189K,2022ApJ...925..126L,ZAKHAROV20141108,Zakharov2018}. \cite{2017PhRvL.118u1101H} studied the scale of modified gravity (yukawa correction to Newtonian potential) near the GC black hole through 19 years of observation of two short period stars S0-2 and S0-38. The scale was reported as $\lambda \ge 150$ au . After detection of the pericentre shift of S0-2 \citep{2020A&A...636L...5G}, \cite{2021PhRvD.104j1502D} reported constraint on the scale of $f(R)$ modified gravity as $\lambda \ge 6300$ au. In this work, first we estimate pericentre shift of stellar orbit in $R^2$ gravity and investigate the deviation from GR prediction. Next, we try to extract the scale of $f(R)$ modified gravity in the context of screening and by using the available bounds on Parametrised Post Newtonian (PPN) parameters near the GC black hole \citep{2020A&A...636L...5G}. For stellar orbits we choose semi-major axes having lower bound imposed by Gravitational Wave (GW) time constraint \citep{2009ApJ...693L..35M,2021ApJ...909..189K} and upper bound imposed by the orbit of S0-2 ($a=1000$ au). We also consider the newly discovered compact orbit of the star S4716 ($a = 407$ au, $P=4$ yrs) \citep{2022ApJ...933...49P}.

\par The paper is organized as follows. Section \ref{b} presents pericentre shift in $f(R)=R^2$ gravity and highlights the deviation from GR prediction. In section \ref{b1} we extract the background $f(R)$ scalaron field amplitude $\psi_0=f'(R_0)$ for three scalaron masses  $M_{\psi}=10^{-22}$ eV, $10^{-19}$ eV and $10^{-16}$ eV. We also identify the scalarons which are unscreened or screened within the considered bound of semi-major axis. In section \ref{b2} we highlight the possibility of detection of pericentre shift through astrometric capabilities of existing and upcoming telescopes. Section \ref{c} presents results and discussions.

\section{Pericentre Shift in $f(R)$ Gravity}\label{b}
The power law gravity, $f(R) \propto R^2$ was considered by \cite{1980PhLB...91...99S} to generate primordial cosmological inflation. It represents higher curvature (high energy) modification to GR in the primordial universe. \cite{2016GrCo...22...71K} demonstrated that power law gravity ($f(R) \propto R^n$, $n \geq 2$) naturally appears due to curvature correction to quantum vacuum fluctuations near a black hole. These corrections are found to be precursor of Yukawa modification to gravitational potential in the weak field regime \citep{Kalita_2018,2020ApJ...893...31K} which has been extensively used to constrain modified gravity near the GC black hole \citep{2017PhRvL.118u1101H,2021PhRvD.104j1502D}. Therefore, testing the effect of power law gravity near the GC black hole is important for using this laboratory for understanding the cosmological problems. Knowing the gravitational potential in the weak field limit is important for estimation of pericentre shift of the orbit of a test particle around a massive object. The weak field limit of power law gravity $f(R)=f_oR^n$ is expressed by the gravitational potential as \citep{2012PhRvD..85l4004B,PhysRevD.72.103005}

\begin{equation}\label{eq:1}
\Phi(r)=-\frac{GM}{2r}\left[ 1+ \left( \frac{r}{r_c} \right)^\delta \right]   .
\end{equation}

Here $r_c$ is an arbitrary parameter related to scale of the system. $\delta$ is a universal parameter expressed as function of $n$ which makes it model dependent. It is expressed as \citep{2012PhRvD..85l4004B}  

\begin{equation}\label{eq:2}
\delta=\frac{12n^2-7n-1-\sqrt{36n^4+12n^3-83n^2+50n+1}}{6n^2-4n+2}   .
\end{equation}

For n=1 and $\delta=0$, $\phi(r)$ reduces to the Newtonian potential as
\begin{equation}\label{eq:3}
\Phi_N(r)=-\frac{GM}{r}   .
\end{equation}

To calculate pericentre shift in $f(R)$ theory the form of perturbing potential is identified by subtracting its Newtonian counterpart as
\begin{equation}
V(r)=\Phi(r)-\Phi_N(r)   ,
\end{equation}
which gives

\begin{equation}\label{eq:4}
V(r)=-\frac{GM}{2r} \left[ \left( \frac{r}{r_c} \right)^\delta-1 \right]   .
\end{equation}

Equation (\ref{eq:4}) is used to calculate pericentre shift using equation (30) of  \cite{PhysRevD.75.082001} as

\begin{equation}\label{eq:5}
\left( \delta \phi \right)_{prec}=\frac{-2L}{GMe^2} \int_{-1}^{1} \frac{zdz}{\sqrt{1-z^2}} \frac{dV(z)}{dz}   ,
\end{equation}

Here, $r$ in (\ref{eq:4}) and $z$ in (\ref{eq:5}) are related as

\begin{equation}
r=\frac{L}{1+ez}   ,
\end{equation} 

and $L$ is the semilatus rectum of the ellipse with semi major axis $a$ and eccentricity $e$ expressed as

\begin{equation}
L=a \left( 1-e^2 \right)   .
\end{equation}

\cite{2012PhRvD..85l4004B} calculated analytical solution of equation (\ref{eq:5}) as

\begin{equation}\label{eq:6}
\left( \delta \phi \right)_{prec}^{f(R)}=\frac{\pi}{2}\delta \left( \delta -1 \right) \left( \frac{a \left(1-e^2 \right)}{r_c} \right)^\delta \times  _2F_1 \left( \frac{\delta + 1}{2},\frac{\delta + 2}{2};2;e^2 \right)   ,
\end{equation}

where  $_2F_1$ is hypergeometric function. It is a function of $\delta$ and hence model dependent. The value of $ r_c$ is chosen to be $100$ au as it gives the maximum allowed range for $\delta$ \citep{2012PhRvD..85l4004B}. The pericentre shift given by (\ref{eq:6}) is considered for both Schwarzschild and Kerr metric exterior to the GC black hole. The $R^2$ model $(n=2)$ taken into account for this work corresponds to $\delta=0.6666$. The difference of pericentre shift between $f(R)$ and Schwarzschild black hole ($(\delta \phi)_{prec}^{Sch}-(\delta \phi)_{prec}^{f(R)}$) and $f(R)$ and Kerr black hole ($(\delta \phi)_{prec}^{Kerr}-(\delta \phi)_{prec}^{f(R)}$), given by this model are plotted against semi-major axis ($a$) for eccentricities $e=0.1$ and $0.9$. The lower bound on semi-major axis for the eccentricities taken into consideration are calculated from the time scale of Gravitational Wave (GW) emission from S-star - GC black hole binary. The time scale of GW emission for a small black hole moving around the GC black hole has been calculated by \cite{2009ApJ...705..361G}. \cite{2021ApJ...909..189K} modified this result for S-star - GC black hole binary system for stellar mass of $10M_\odot$. It is expressed as

\begin{equation}\label{eq:7}
t_{GW}=\frac{5c^5}{256f(e)G^3} \frac{a^4}{\mu(M_{BH}+M_\star)^2}   ,
\end{equation}

where
\begin{equation}
f(e)=\left( 1-e^2 \right)^{-\frac{7}{2}} \left( 1+ \frac{73}{24}e^2+\frac{37}{96}e^4 \right)   ,
\end{equation}
and
\begin{equation}
\mu=\frac{M_{BH}M_\star}{(M_{BH}+M_\star)}   .
\end{equation}
Here, $ M_\star$ is the mass of an S-star and $M_{BH}$ is the mass of GC black hole. Studies on the nuclear star cluster near the GC black hole give a bound on age of these stars within $(6-400)$ Myr \citep{2010RvMP...82.3121G}. This suggests that these stars must have survived for at least $6$ Myr before their orbits undergo rapid evolution through GW emission. Using this assumption the minimum bound on $a$ is calculated from (\ref{eq:7}) for two eccentricities $0.1$ and $0.9$. They are presented in Table \ref{tab1}.

\begin{table}[hptb]
\begin{center}
\caption{Lower bound on semi-major axis from GW time scale constraint.}
{\begin{tabular}{@{}cc@{}} \hline \hline
Eccentricity($e$) & Minimum bound on Semi-major axis($a_{min}$) \\ \hline
0.1             & 7.71 au                             \\
0.9             & 45.40 au\\ \hline
\end{tabular} \label{tab1}}
\end{center}
\end{table}

\subsection{Schwarzschild black hole}
The pericentre shift due to Schwarzschild solution is expressed as
\begin{equation}\label{eq:8}
\left( \delta \phi \right)_{prec}^{Sch}=\frac{6\pi GM_{BH}}{a \left( 1 - e^2 \right)c^2}   .
\end{equation}
Here, $M_{BH}$ is the mass of the black hole ($ \approx 4.261 \times 10^6\ M_{\odot}$).
The deviation from Schwarzschild pericentre shift is expressed as
\begin{equation}\label{eq:9}
\left( \delta \phi \right)_{prec}^{Sch}-\left( \delta \phi \right)_{prec}^{f(R)}=\frac{6\pi G M_{BH}}{a(1-e^2)c^2}-\frac{\pi}{2}\delta \left( \delta -1 \right) \left( \frac{a \left(1-e^2 \right)}{r_c} \right)^\delta \times  _2F_1 \left( \frac{\delta + 1}{2},\frac{\delta + 2}{2};2;e^2 \right)   .
\end{equation}

\begin{figure}
    \centering
    \begin{subfigure}[b]{0.45\textwidth}
        \includegraphics[width=\textwidth]{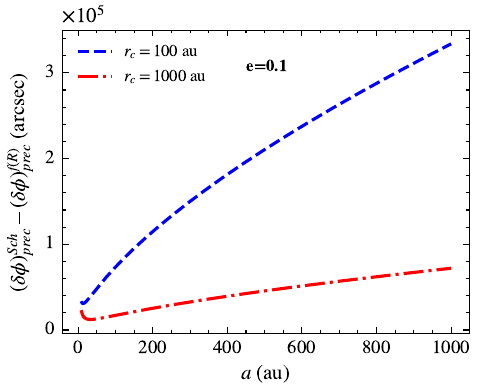}
    \end{subfigure}
    \begin{subfigure}[b]{0.45\textwidth}
        \includegraphics[width=\textwidth]{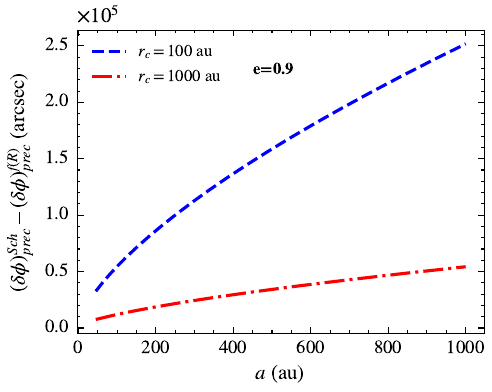}
    \end{subfigure}
    \caption{Variation of difference between Schwarzschild and $f(R)$ pericentre shift against semi major axis ($a$) for eccentricity, $e=0.1$ (left figure) and $e=0.9$ (right figure)}
    \label{fig1}
\end{figure}

The deviation in pericentre shift expressed by equation (\ref{eq:9}) is plotted against Semi-major axis, $a$ for two $r_c$ values, $100$ au and $1000$ au for $R^2$ $(n=2)$ model and have been presented in Fig.~\ref{fig1}.

\subsection{Kerr black hole}
The pericentre shift derived from Kerr solution is expressed as,\citep{2011GReGr..43..977D}
\begin{equation}\label{eq:10}
\left(\delta \phi \right)_{prec}^{Kerr}=\frac{6\pi GM_{BH}}{a \left( 1 - e^2 \right)c^2}+\frac{8\pi a' M_{BH}^{\frac{1}{2}}G^{\frac{3}{2}}}{a^{\frac{3}{2}}(1-e^2)^{\frac{3}{2}}c^3} .
\end{equation}
Here, $a'=J/M_{BH}c$ is the Kerr parameter and is expressed in terms of spin of the black hole as $\chi=a'/m = Jc/G{M_{BH}}^2$ (where, $m=G{M_{BH}}/c^2)$ such that (\ref{eq:10}) takes the form,
\begin{equation}\label{eq:11}
\left( \delta \phi \right)_{prec}^{Kerr}=\frac{6\pi}{a(1-e^2)}\left( \frac{GM_{BH}}{c^2} \right)+\frac{8\pi\chi}{[a(1-e^2)]^{\frac{3}{2}}} \left( \frac{GM_{BH}}{c^2}\right)^{\frac{3}{2}}  .
\end{equation}
The first term of (\ref{eq:11}) is the non-rotating Schwarzschild component and the second term is the contribution due to the spin of the black hole. The difference in pericentre shift for two theories is plotted against semi-major axis ($a$) for two values of black hole spin $\chi=0.1$ and $0.9$ for each eccentricity value. The value $\chi=0.1$ is taken from the observational constraints on orientation of stellar orbits near the GC black hole as reported by \cite{Fragione_2020}. $\chi=0.9$ has been chosen freely so as to have a large contribution from spin induced pericentre shift. In this case the deviation is expressed as

\begin{multline}\label{eq:12}
\left( \delta \phi \right)_{prec}^{Kerr}-\left( \delta \phi \right)_{prec}^{f(R)}=\frac{6\pi}{a(1-e^2)}\left( \frac{GM_{BH}}{c^2} \right)+\frac{8\pi\chi}{[a(1-e^2)]^{\frac{3}{2}}} \left( \frac{GM_{BH}}{c^2}\right)^{\frac{3}{2}}\\ -\frac{\pi}{2}\delta \left( \delta -1 \right) \left( \frac{a \left(1-e^2 \right)}{r_c} \right)^\delta \times  _2F_1 \left( \frac{\delta + 1}{2},\frac{\delta + 2}{2};2;e^2 \right)   ,
\end{multline}

\begin{figure}
    \centering
    \begin{subfigure}[b]{0.45\textwidth}
        \includegraphics[width=\textwidth]{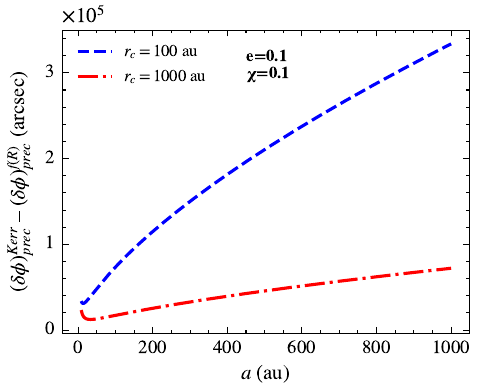}
    \end{subfigure}
    \begin{subfigure}[b]{0.45\textwidth}
        \includegraphics[width=\textwidth]{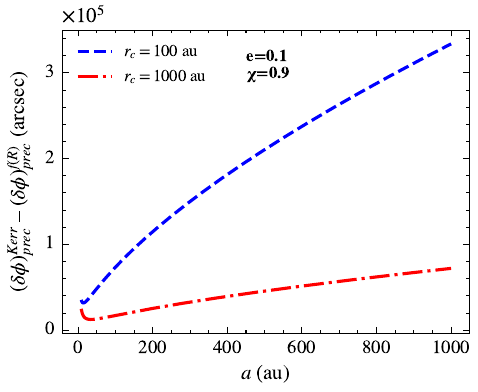}
    \end{subfigure}\\
    \begin{subfigure}[b]{0.45\textwidth}
        \includegraphics[width=\textwidth]{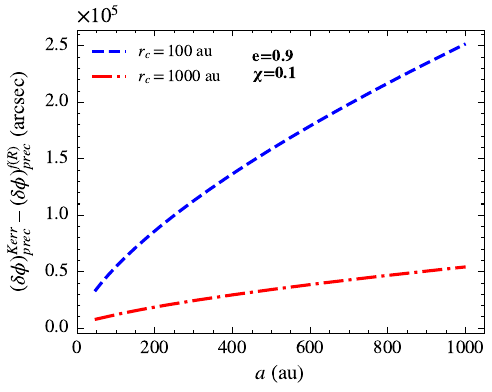}
    \end{subfigure}
    \begin{subfigure}[b]{0.45\textwidth}
        \includegraphics[width=\textwidth]{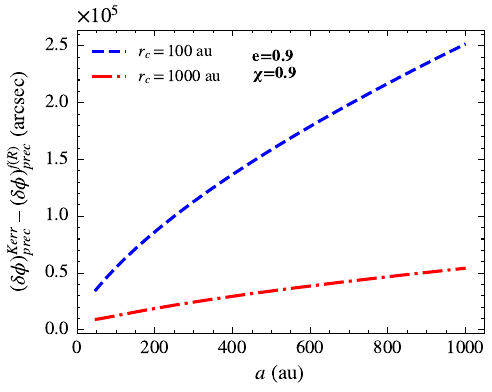}
    \end{subfigure}
    \caption{Variation of difference between Kerr and $f(R)$ pericentre shift against semi major axis ($a$) for eccentricity, $e=0.1$ \& spin, $\chi=0.1$ (top left figure); $e=0.1$ \& $\chi=0.9$ (top right figure); $e=0.9$ \& $\chi=0.1$ (bottom left figure) and $e=0.9$ \& $\chi=0.9$ (bottom right figure)}
    \label{fig2}
\end{figure}

The variation of deviation is shown in Fig.\ref{fig2}. 

It has been observed that the difference between $f(R)$ and GR pericentre shift goes up with semi-major axis for both Kerr black holes and Schwarzschild black holes. This growth becomes prominent for $r_c=100$ au. It is found that the pericentre shift due to $f(R)$ gravity approaches to that in GR only for very compact orbits ($a\approx$ few tens of au). The effect of modified gravity gets suppressed by GR towards wider orbits. We test effect of modified gravity in a model independent manner in the next section.

\section{Parameter Extraction and Screening of $f(R)$ gravity}\label{b1}

The weak field limit of $f(R)$ gravity presents a Yukawa correction to the Newtonian potential ($-GM/r$) in the form $-\frac{GM}{r}\frac{1}{3\psi_{0}}e^{-M_{\psi} r}$ , where $1/3\psi_{0}=\alpha$ is the Yukawa coupling strength generated by the dimensionless scalaron field amplitude $\psi_{0}$ \citep{2020ApJ...893...31K} and $M_{\psi}$ is the mass of the scalaron field. The general formula for rate of pericentre shift in f(R) gravity is expressed as (see \cite{2014RAA....14..139L} for pericentre shift due to Yukawa correction)

\begin{equation}\label{eq:13}
\left( \delta \dot{\phi}\right)_{prec}^{SC}=\alpha\frac{na}{e\lambda}\sqrt{1-e^2}\exp{\left(-\frac{a}{\lambda}\right)}I_1\left( \frac{ae}{\lambda} \right)   ,
\end{equation}

where, $n=\frac{2\pi}{P}$ (here $P$ is orbital period), $\lambda=M_{\psi}^{-1}$ is a scale length corresponding to the scalaron force and $I_1(x)$ is the modified Bessel function of first kind of index 1. Equation (\ref{eq:13}) as adopted by \cite{2020ApJ...893...31K} can be written for pericentre shift angle as,

\begin{equation}\label{eq:14}
\left( \delta \phi \right)_{prec}^{SC}=\frac{2\pi}{3\psi_{0}}\left( \frac{M_{\psi}a}{e} \right)\sqrt{1-e^2}\exp{\left(-M_{\psi}a\right)}I_1(M_{\psi}ae)   .
\end{equation}

Equation (\ref{eq:14}) is equated to pericentre shift angle in Parametrised Post Newtonian (PPN) formalism \citep{2001LRR.....4....4W},
\begin{equation}\label{eq:15}
\left( \delta \phi \right)_{prec}^{PPN}=\frac{1}{3}(2+2\gamma-\beta)\frac{6\pi GM_{BH}}{c^2a(1-e^2)}   ,
\end{equation}

where, $\gamma$ and $\beta$ are the PPN parameters whose values near the GC black hole are adopted from Ref. \cite{2020A&A...636L...5G}. Therefore, using (\ref{eq:14}) and (\ref{eq:15}) a relation between scalaron field $\psi_{0}$ and PPN parameters $\gamma$ and $\beta$ is established as

\begin{equation}\label{eq:16}
\frac{2\pi}{3\psi_{0}}\left( \frac{M_{\psi}a}{e} \right)\sqrt{1-e^2}\exp{\left(-M_{\psi}a\right)}I_1(M_{\psi}ae)=\frac{1}{3}(2+2\gamma-\beta)\frac{6\pi GM_{BH}}{c^2a(1-e^2)}   .
\end{equation}

In this section we extract the allowed ranges of the scalaron field $\psi_0$ and hence Yukawa coupling strength $\alpha$ with the help of PPN parameters ($\gamma,\beta$) constrained by \cite{2020A&A...636L...5G} through the measurement of pericentre shift of S0-2. The scalaron masses are chosen as $10^{-22}$ eV, $10^{-19}$ eV and $10^{-16}$ eV (with the conversion $1$ au$^{-1}=8.25 \times 10^{-18}$ eV) (see \cite{2020ApJ...893...31K,2021ApJ...909..189K,2022ApJ...925..126L} for details of mass ranges). The scalaron field $\psi_0$ is plotted with respect to $\gamma \ \& \ \beta$ bounded by the error bars of their constraints. These constraints are \citep{2020A&A...636L...5G}

$$\gamma=1.18 \pm 0.34 \ ,$$  
$$\beta=1.05 \pm 0.11 \ $$

\begin{figure}
    \centering
    \begin{subfigure}[b]{0.24\textwidth}
        \includegraphics[width=\textwidth]{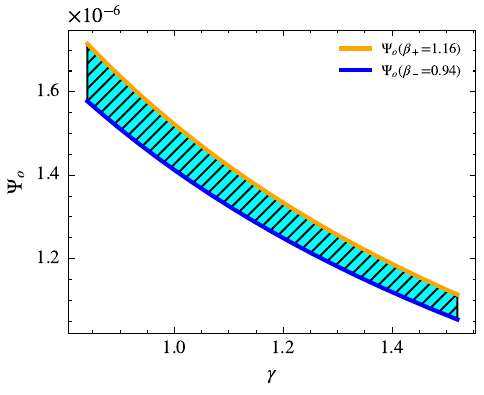}
    \end{subfigure}
    \begin{subfigure}[b]{0.24\textwidth}
        \includegraphics[width=\textwidth]{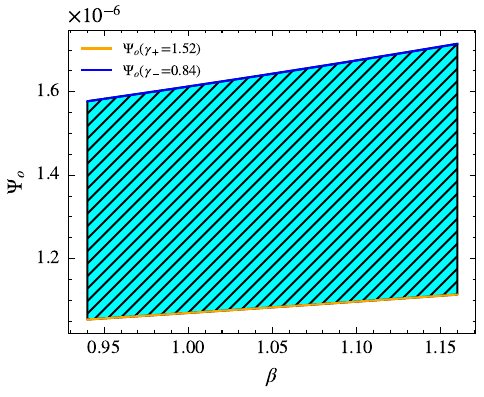}
    \end{subfigure}
    \begin{subfigure}[b]{0.24\textwidth}
        \includegraphics[width=\textwidth]{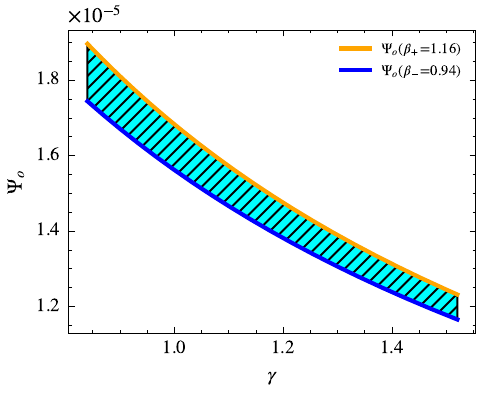}
    \end{subfigure}
    \begin{subfigure}[b]{0.24\textwidth}
        \includegraphics[width=\textwidth]{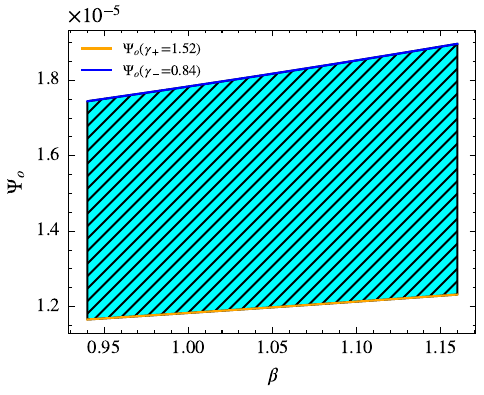}
    \end{subfigure}\\
    \begin{subfigure}[b]{0.24\textwidth}
        \includegraphics[width=\textwidth]{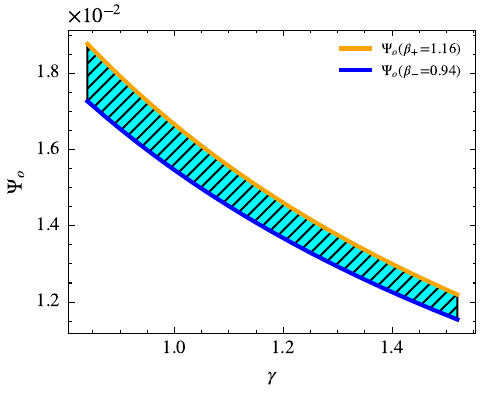}
    \end{subfigure}
    \begin{subfigure}[b]{0.24\textwidth}
        \includegraphics[width=\textwidth]{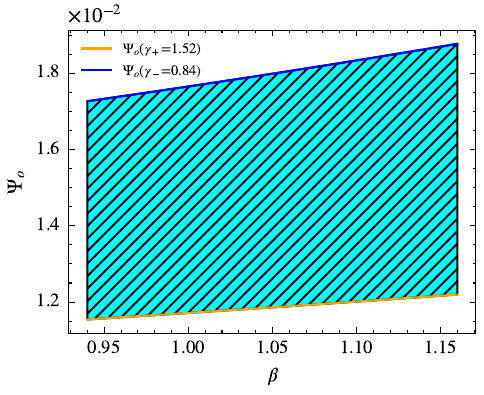}
    \end{subfigure}
    \begin{subfigure}[b]{0.24\textwidth}
        \includegraphics[width=\textwidth]{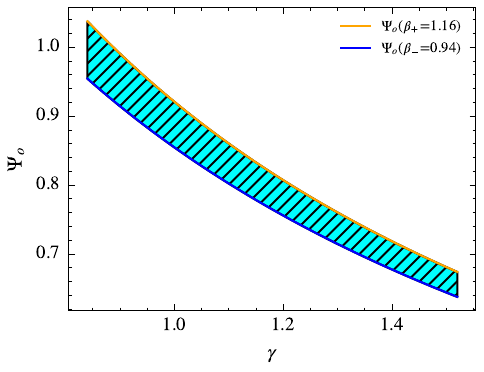}
    \end{subfigure}
    \begin{subfigure}[b]{0.24\textwidth}
        \includegraphics[width=\textwidth]{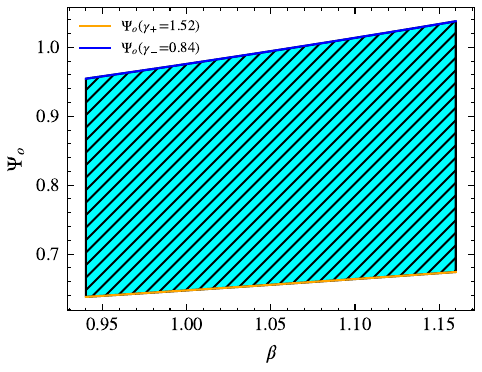}
    \end{subfigure}\\
    \begin{subfigure}[b]{0.24\textwidth}
        \includegraphics[width=\textwidth]{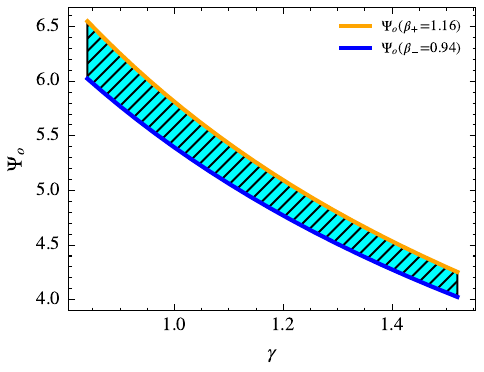}
    \end{subfigure}
    \begin{subfigure}[b]{0.24\textwidth}
        \includegraphics[width=\textwidth]{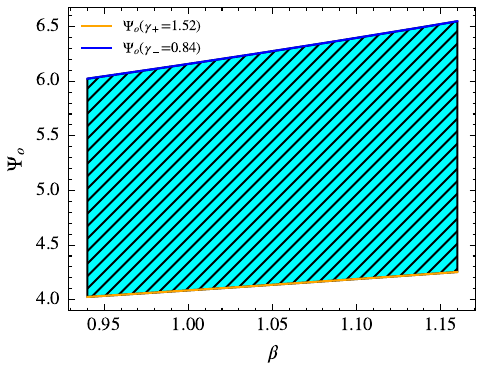}
    \end{subfigure}
    \begin{subfigure}[b]{0.24\textwidth}
        \includegraphics[width=\textwidth]{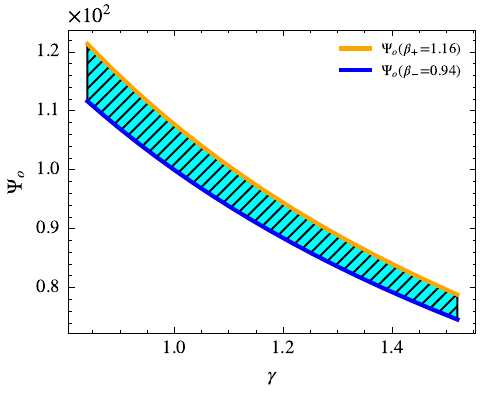}
    \end{subfigure}
    \begin{subfigure}[b]{0.24\textwidth}
        \includegraphics[width=\textwidth]{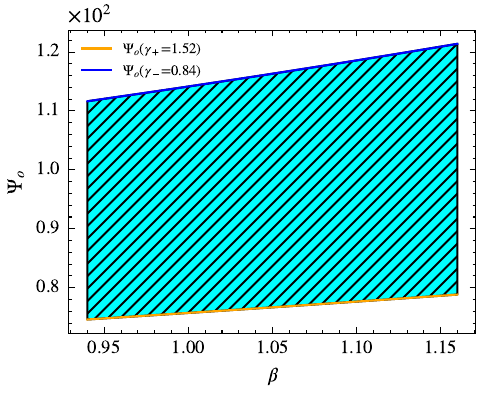}
    \end{subfigure}\\
    \begin{subfigure}[b]{0.24\textwidth}
        \includegraphics[width=\textwidth]{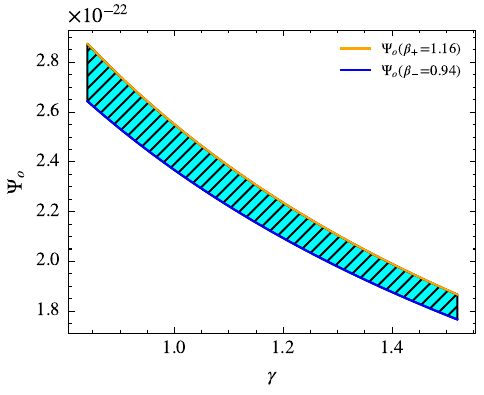}
    \end{subfigure}
    \begin{subfigure}[b]{0.24\textwidth}
        \includegraphics[width=\textwidth]{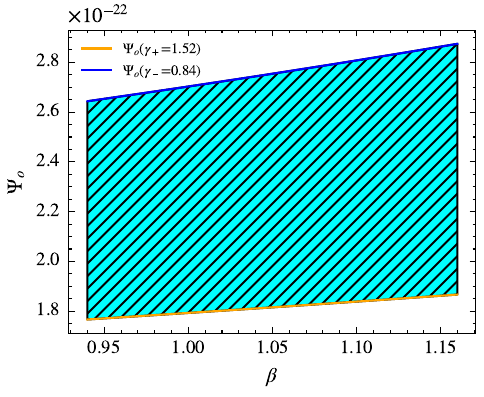}
    \end{subfigure}
    \caption{Allowed range of $\psi_0$ for $M_{\psi}=10^{-22}$ eV at $a=45$ au (first and second figures from the left in top row), $a=100$ au (third and fourth figures from the left in top row) and $a=1000$ au (first and second figures from the left in second row); for $M_\psi=10^{-19}$ eV at $a=45$ au (third and fourth figures from the left in second row), $a=100$ au (first and second figures from the left in third row) and $a=1000$ au (third and fourth figures from the left in third row) and for $M_\psi=10^{-16}$ eV at $a=45$ au (first and second figures from the left in bottom row).}
    \label{fig3}
\end{figure}

These bounds are relatively poor in comparison to the stringent bounds realised through solar system tests \citep{2003Natur.425..374B,1979ApJ...234L.219R,1990grg..conf..313S}. However, these bounds have been achieved in a new environment of the black hole and hence can be employed to understand modified gravity. For a given scalaron mass the variations are studied at semi-major axes $45$ au, $100$ au and $1000$ au by taking into account the two masses $10^{-22}$ eV and $10^{-19}$ eV \footnote{For $M_{\psi}=10^{-16}$ eV plots for $100$ au and $1000$ au are not shown as the exponential in left hand side of (\ref{eq:16}) is vanishingly small due to large negative powers}. These variations are shown in Fig. \ref{fig3}. Meaning of these allowed regions of $\psi_{0}$ in ($\psi_{0} , \gamma , \beta$) space is further illustrated by the method of screening of $f(R)$ gravity (see next section). For each value of scalaron mass the extracted ranges of the scalaron field and Yukawa coupling are displayed in Table \ref{tab2}. It has been calculated from (\ref{eq:16}) by considering the above constraints on $\gamma, \beta$.

\begin{table}[hptb]
\caption{Bounds on $\psi_{0}$ and $\alpha$ for different scalaron masses. The ranges of $\psi_0$ and $\alpha$ are due to uncertainities in $\gamma$ and $\beta$.}
\centering
{\begin{tabular}{@{}cccc@{}} \hline \hline
{\shortstack[m]{Scalaron mass\\($M_{\psi}$)}} &{\shortstack[m]{Semi-major axis\\($a$)}} & {\shortstack[m]{Bounds on $\psi_{0}$\\(Min - Max)}} & {\shortstack[m]{Bounds on $\alpha$\\(Min - Max)}} \\ \hline
\multirow{3}{*}{}{$10^{-22}$ eV} & \hphantom{0}45 au & \hphantom{0}$1.05 \times 10^{-6} - 1.71 \times 10^{-6}$ & \hphantom{0}$1.94 \times 10^5 - 3.17 \times 10^5$ \\
& \hphantom{0}100 au & \hphantom{0}$1.16 \times 10^{-5} - 1.89 \times 10^{-5}$ & \hphantom{0}$1.76 \times 10^4 - 2.87 \times 10^4$\\
& \hphantom{0}1000 au & \hphantom{0}$0.01154 - 0.01877$ & \hphantom{0}$17.75 - 28.88$\\ \hline
$10^{-19}$ eV & \hphantom{0}45 au & \hphantom{0}$0.6378 - 1.0378$ & \hphantom{0}$0.32 - 0.52$\\
& \hphantom{0}100 au & \hphantom{0}$4.0253 - 6.5491 $ & \hphantom{0}$0.05 - 0.08$\\
& \hphantom{0}1000 au & \hphantom{0}$74.6127 - 121.3938$ & \hphantom{0}$2.74 \times 10^{-3} - 4.46 \times 10^{-3}$ \\ \hline
$10^{-16}$ eV & \hphantom{0}45 au & \hphantom{0}$1.76 \times 10^{-22} - 2.87 \times 10^{-22} $ & \hphantom{0}$1.16 \times 10^{21} - 1.89 \times 10^{21} $\\ \hline
\end{tabular} \label{tab2}}
\end{table}

\par The scalaron field is screened if $\psi_{0}/\phi < 1$ and is unscreened if $\psi_{0}/\phi > 1$ \citep{2012A&G....53d..37L} where $\phi$ is the dimensionless gravitational potential expressed as

\begin{equation}
\phi=\frac{GM_{BH}}{c^2r}   .
\end{equation}

\begin{figure}
    \centering
    \begin{subfigure}[b]{0.24\textwidth}
        \includegraphics[width=\textwidth]{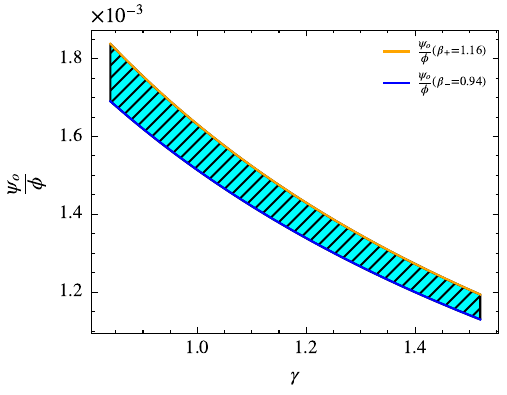}
    \end{subfigure}
    \begin{subfigure}[b]{0.24\textwidth}
        \includegraphics[width=\textwidth]{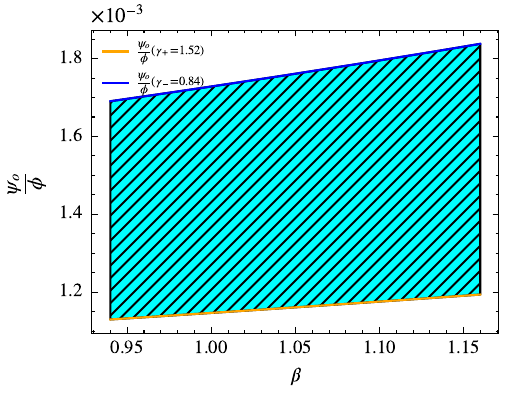}
    \end{subfigure}
    \begin{subfigure}[b]{0.24\textwidth}
        \includegraphics[width=\textwidth]{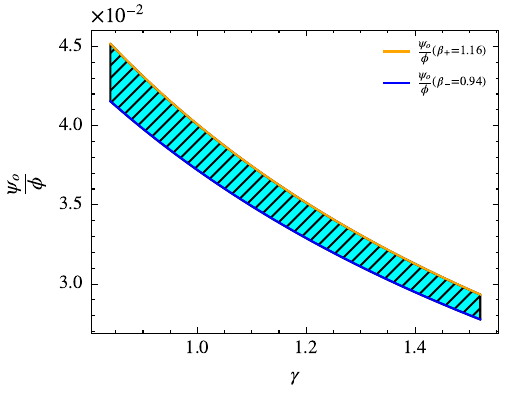}
    \end{subfigure}
    \begin{subfigure}[b]{0.24\textwidth}
        \includegraphics[width=\textwidth]{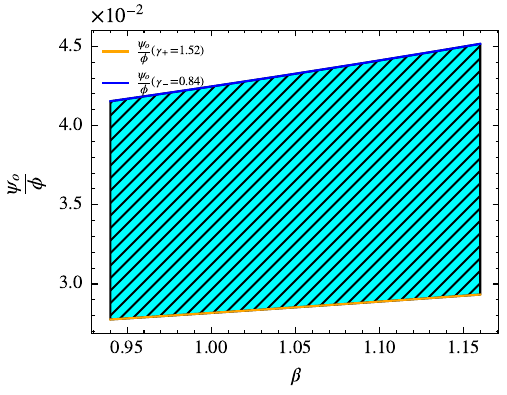}
    \end{subfigure}\\
    \begin{subfigure}[b]{0.24\textwidth}
        \includegraphics[width=\textwidth]{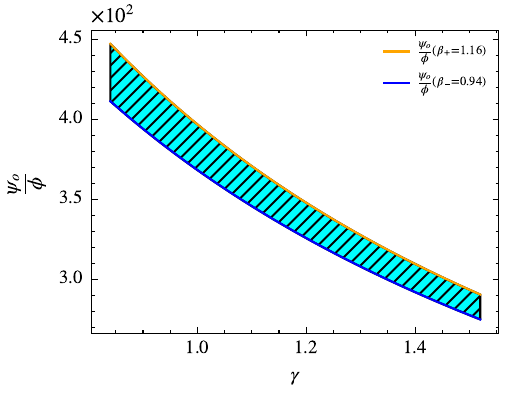}
    \end{subfigure}
    \begin{subfigure}[b]{0.24\textwidth}
        \includegraphics[width=\textwidth]{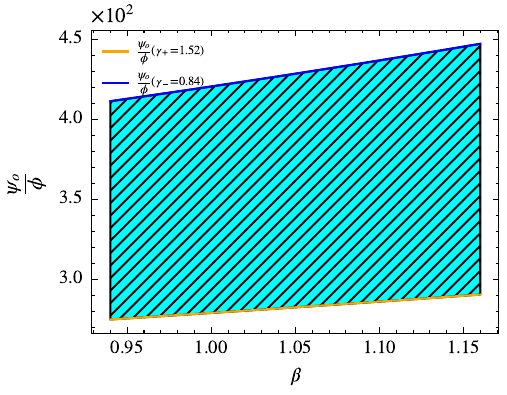}
    \end{subfigure}
    \begin{subfigure}[b]{0.24\textwidth}
        \includegraphics[width=\textwidth]{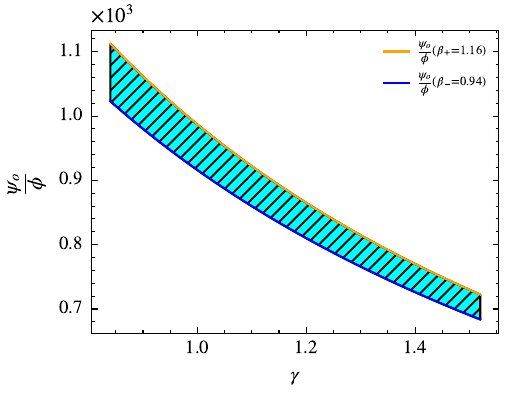}
    \end{subfigure}
    \begin{subfigure}[b]{0.24\textwidth}
        \includegraphics[width=\textwidth]{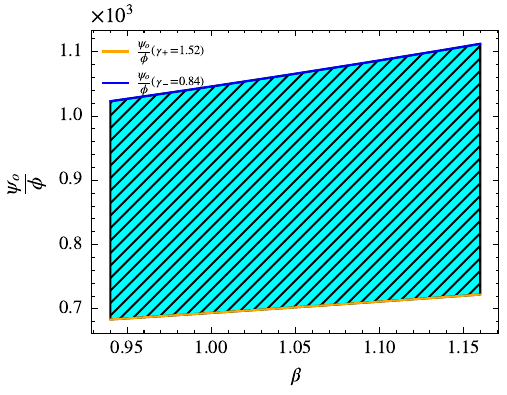}
    \end{subfigure}\\
    \begin{subfigure}[b]{0.24\textwidth}
        \includegraphics[width=\textwidth]{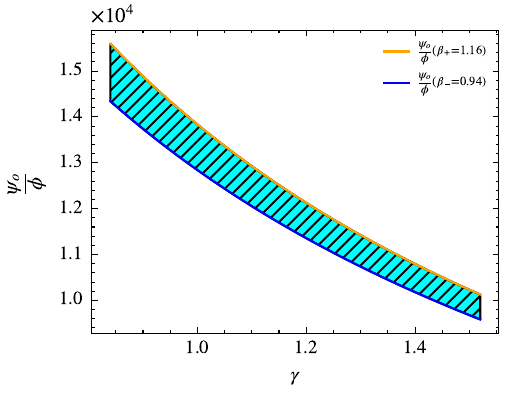}
    \end{subfigure}
    \begin{subfigure}[b]{0.24\textwidth}
        \includegraphics[width=\textwidth]{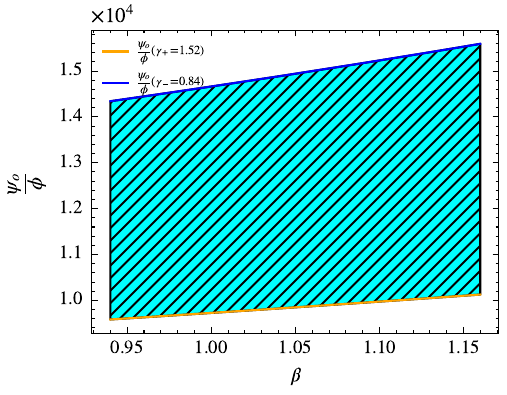}
    \end{subfigure}
    \begin{subfigure}[b]{0.24\textwidth}
        \includegraphics[width=\textwidth]{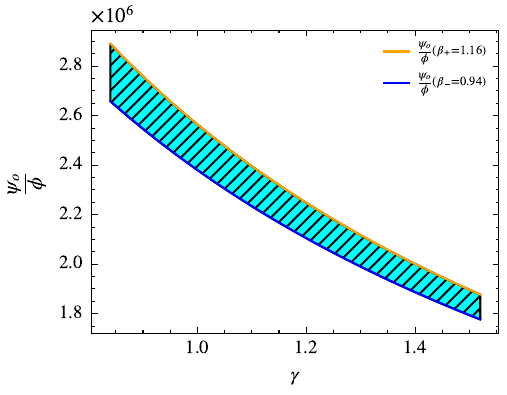}
    \end{subfigure}
    \begin{subfigure}[b]{0.24\textwidth}
        \includegraphics[width=\textwidth]{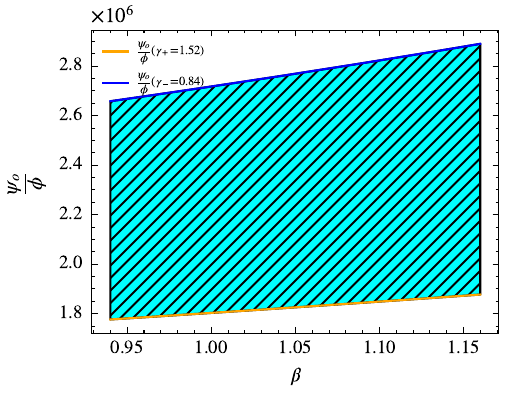}
    \end{subfigure}\\
    \begin{subfigure}[b]{0.24\textwidth}
        \includegraphics[width=\textwidth]{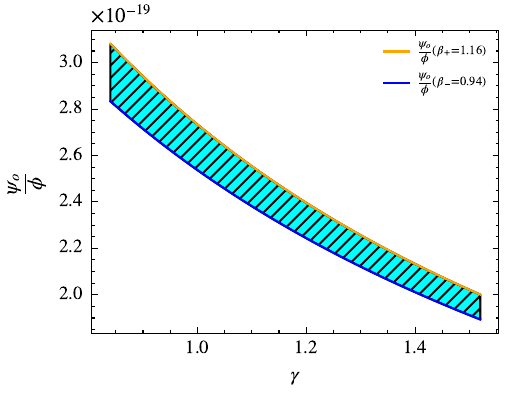}
    \end{subfigure}
    \begin{subfigure}[b]{0.24\textwidth}
        \includegraphics[width=\textwidth]{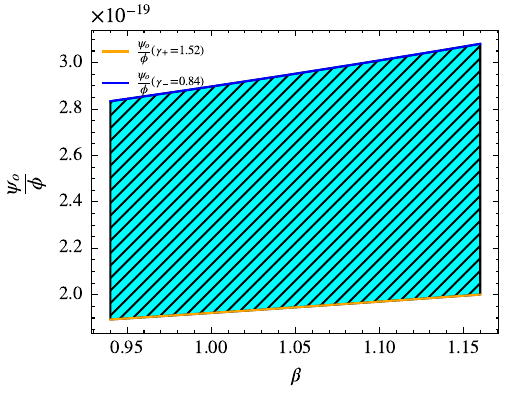}
    \end{subfigure}
    \caption{Range of $\psi_0/\phi$ for $M_\psi=10^{-22}$ eV at $a=45$ au (first and second figures from the left in top row), $a=100$ au (third and fourth figures from the left in top row) and $a=1000$ au (first and second figures from the left in second row); for $M_\psi=10^{-19}$ eV at $a=45$ au (third and fourth figures from the left in second row), $a=100$ au (first and second figures from the left in third row) and $a=1000$ au (third and fourth figures from the left in third row) and for $M_\psi=10^{-16}$ eV at $a=45$ au (first and second figures from the left in bottom row).}
    \label{fig4}
\end{figure}

In order to examine screening or unscreening of $f(R)$ gravity we plot $\psi_{0}/\phi$ against variations in $\gamma$ and $\beta$. These variations are shown in Fig.\ref{fig4}. Table \ref{tab3} highlights the locations near the GC black hole where $f(R)$ gravity is screened or unscreened. From the table it is evident that light scalarons ($10^{-22}$ eV) are unscreened at $1000$ au. Slightly heavier scalarons ($10^{-19}$ eV), however, remain unscreened at all the semi-major axes.

\begin{table}[hptb]
\caption{$\frac{\psi_{0}}{\phi}$ for different scalaron masses.}
\centering
{\begin{tabular}{@{}cccc@{}} \hline \hline
{\shortstack[m]{Scalaron mass\\($M_{\psi}$)}} & {\shortstack[m]{Semi-major axis\\($a$)}} & {\shortstack[m]{Eccentricity\\($e$)}} & {\shortstack[m]{Screening\\$\frac{\psi_{0}}{\phi}$}} \\ \hline
\multirow{3}{*}{}{$10^{-22}$ eV} & \hphantom{0}$45$ au &  \hphantom{0}$0.9$ & \hphantom{0}$<1$ ,$\ \ \ $Screened \\
& \hphantom{0}$100$ au & \hphantom{0}$0.9$ & \hphantom{0}$<1$ ,$\ \ \ $Screened\\
& \hphantom{0}$1000$ au & \hphantom{0}$0.9$ & \hphantom{0}$>1$ ,Unscreened\\ \hline
$10^{-19}$ eV & \hphantom{0}$45$ au &  \hphantom{0}$0.9$ & \hphantom{0}$>1$ ,Unscreened\\
& \hphantom{0}$100$ au & \hphantom{0}$0.9$ & \hphantom{0}$>1$ ,Unscreened\\
& \hphantom{0}$1000$ au & \hphantom{0}$0.9$ & \hphantom{0}$>1$ ,Unscreened\\ \hline
$10^{-16}$ eV & \hphantom{0}$45$ au & \hphantom{0}$0.9$ & \hphantom{0}$<1$ ,$\ \ \ $Screened\\
& \hphantom{0}$100$ au & \hphantom{0}$0.9$ & \hphantom{0}$<1$ ,$\ \ \ $Screened\\
& \hphantom{0}$1000$ au & \hphantom{0}$0.9$ & \hphantom{0}$<1$ ,$\ \ \ $Screened\\ \hline
\end{tabular} \label{tab3}}
\end{table}

\par
\begin{figure}
    \centering
    \begin{subfigure}[b]{0.24\textwidth}
        \includegraphics[width=\textwidth]{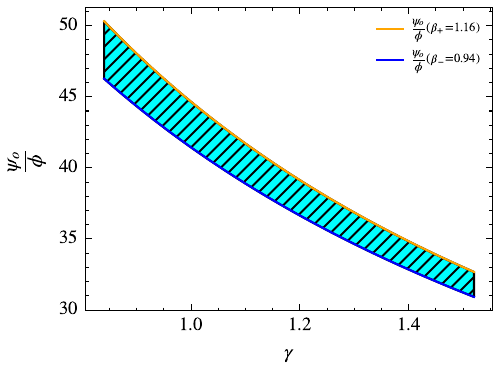}
    \end{subfigure}
    \begin{subfigure}[b]{0.24\textwidth}
        \includegraphics[width=\textwidth]{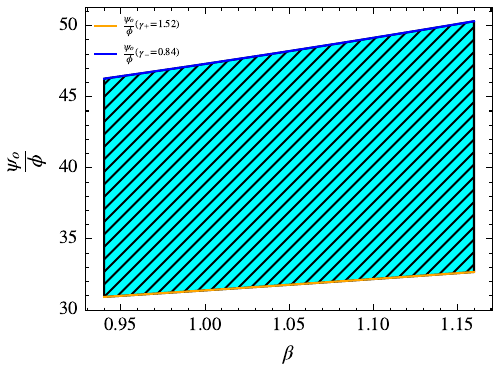}
    \end{subfigure}
    \begin{subfigure}[b]{0.24\textwidth}
        \includegraphics[width=\textwidth]{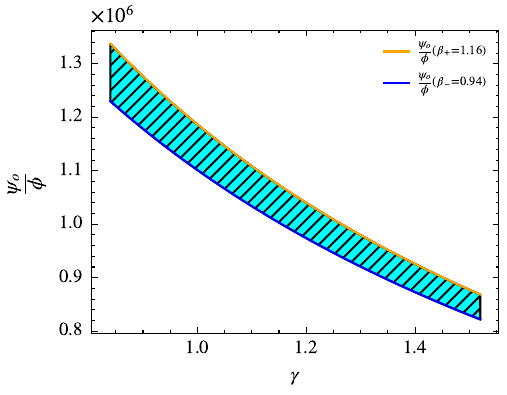}
    \end{subfigure}
    \begin{subfigure}[b]{0.24\textwidth}
        \includegraphics[width=\textwidth]{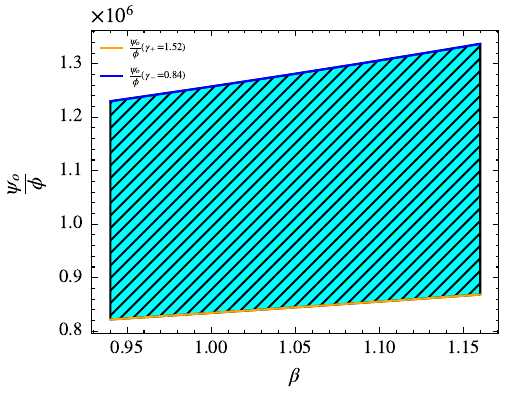}
    \end{subfigure}\\
    \caption{Range of $\psi_0/\phi$ in the location of S4716, for $M_\psi=10^{-22}$ eV (first and second figures from the left) and for $M_\psi=10^{-19}$ eV (third and fourth figures from the left).}
    \label{fig5}
\end{figure}

\begin{figure}
    \centering
    \begin{subfigure}[b]{0.24\textwidth}
        \includegraphics[width=\textwidth]{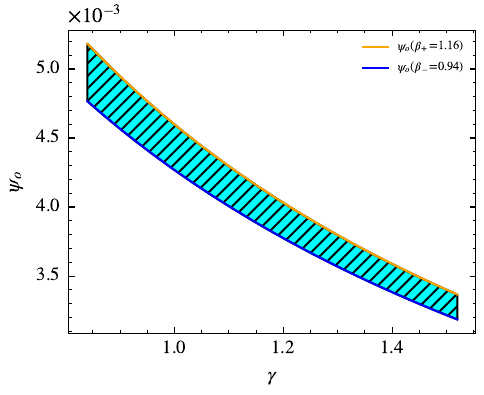}
    \end{subfigure}
    \begin{subfigure}[b]{0.24\textwidth}
        \includegraphics[width=\textwidth]{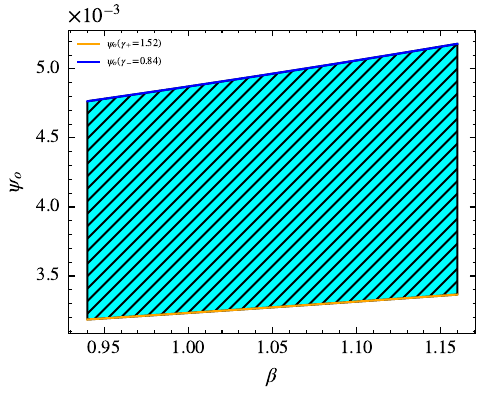}
    \end{subfigure}
    \begin{subfigure}[b]{0.24\textwidth}
        \includegraphics[width=\textwidth]{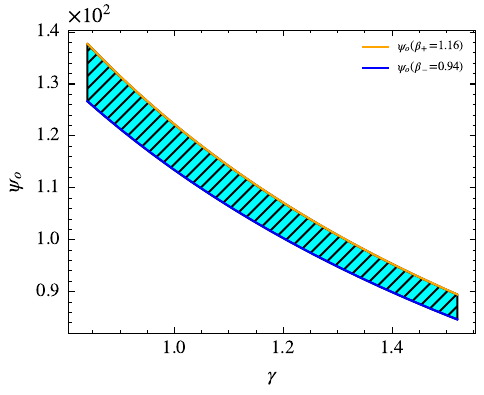}
    \end{subfigure}
    \begin{subfigure}[b]{0.24\textwidth}
        \includegraphics[width=\textwidth]{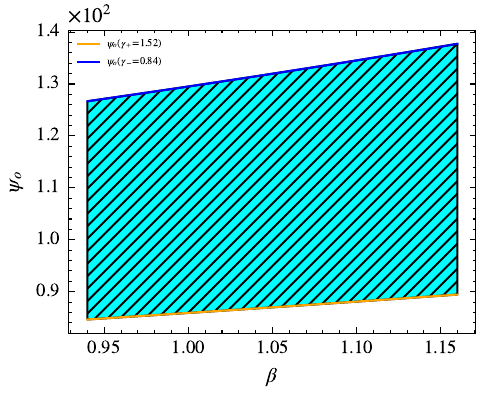}
    \end{subfigure}\\
    \caption{Allowed range of $\psi_0$ in the location of S4716, for $M_\psi=10^{-22}$ eV (first and second figures from the left) and for $M_\psi=10^{-19}$ eV (third and fourth figures from the left).}
    \label{fig6}
\end{figure}

 We examine the screening of $f(R)$ gravity for the newly discovered short period star S4716 ($a=407$ au, $e=0.7$) \citep{2022ApJ...933...49P}. The dimensionless gravitational potential in the location of this star is estimated as $\phi=\frac{GM_{BH}}{c^2 a}$. The variations of $\psi_0/\phi$ for $\gamma$ and $\beta$ are presented in Fig.\ref{fig5}. Table \ref{tab4} presents the screening/unscreening of the scalarons in the orbit of this star\footnote{$M_\psi=10^{-16}$ eV is considered screened as the exponential term in the left hand side of (\ref{eq:16}) vanishes due to large negative powers}. It is seen that the light scalarons ($M_\psi=10^{-22}$ eV) and slightly heavier scalarons ($M_\psi=10^{-19}$ eV) remain unscreened for the star S4716 as well. The heavier scalarons ($M_\psi=10^{-16}$ eV), however, remain completely screened. For the unscreened scalarons near S4716 we extract the allowed range on the scalaron field amplitude ($\psi_0$). We use this range of $\psi_0$ in equation (\ref{eq:14}) to get the allowed range on pericentre shift of S4716 due to these unscreened scalaron masses ($M_\psi=10^{-22}$ eV and $10^{-19}$ eV). The variations are shown in Fig.\ref{fig6}. The range of $\psi_0$ is presented in Table \ref{tab5}.

\begin{table}[hptb]
\caption{Screening of $f(R)$ gravity in the orbit of S4716 ($a=407$ au, $e=0.7$).}
\centering
{\begin{tabular}{@{}cc@{}} \hline \hline
{\shortstack[m]{Scalaron mass ($M_{\psi}$)}} & {\shortstack[m]{Screening ($\frac{\psi_{0}}{\phi}$)}} \\ \hline
$10^{-22}$ eV & \hphantom{0}$>1$ ,Unscreened \\
$10^{-19}$ eV & \hphantom{0}$>1$ ,Unscreened\\ 
$10^{-16}$ eV & \hphantom{0}$<1$ ,$\ \ \ $Screened\\ \hline
\end{tabular} \label{tab4}}
\end{table}

\begin{table}[hptb]
\caption{Bounds on $\psi_{0}$ and pericentre shift for unscreened scalaron masses in the orbit of S4716.}
\centering
{\begin{tabular}{@{}ccccc@{}} \hline \hline
{\shortstack[m]{Scalaron Mass\\($M_\psi$)}} & {\shortstack[m]{Bounds on $\psi_{0}$\\(Min - Max)}} & {\shortstack[m]{Corresponding \\ bounds on $\alpha$}} & {\shortstack[m]{Pericentre Shift\\$(\delta\phi)_{prec}^{SC}$ (arc min)}} \\ \hline
$10^{-22}$ eV & \hphantom{0}$0.0052-0.0032$ &  \hphantom{0}$64.32-104.65$ & \hphantom{0}$11.65-18.95$  \\
$10^{-19}$ eV & \hphantom{0}$137.74-84.66$ & \hphantom{0}$2.42 \times 10^{-3} - 3.94 \times 10^{-3}$ & \hphantom{0}$11.65 - 18.96$\\ \hline
\end{tabular} \label{tab5}}
\end{table}

\par The allowed range of scalaron field amplitude ($\psi_0$) in Table \ref{tab5} is extracted from equation (\ref{eq:16}) and using the bounds on PPN parameters ($\gamma,\beta$) near the GC black hole. We further investigate the validity of this result. We take the Schwarzschild pericentre shift for S4716 from \cite{2022ApJ...933...49P} as $(\delta\phi)_{prec}^{Sch}=14'.8 \pm 1'.3$ and extract $\psi_0$ using the relation (assuming that the scalaron induced pericentre shift is of the order of GR induced shift)

\begin{equation}\label{eq:17}
\frac{2\pi}{3\psi_{0}}\left( \frac{M_{\psi}a}{e} \right)\sqrt{1-e^2}\exp{\left(-M_{\psi}a\right)}I_1(M_{\psi}ae)=14'.8 \pm 1'.3  \ .
\end{equation}

From (\ref{eq:17}) the allowed range of $\psi_0$ for light scalarons ($M_\psi=10^{-22}$ eV) is found to be $0.004 - 0.003$. For slightly heavier scalarons ($M_\psi=10^{-19}$ eV) the allowed range of $\psi_0$ is found to be $118.82 - 99.637$. Thus, our extracted range of $\psi_0$, shown in Table \ref{tab5} is found to be consistent with the values derived from observed pericentre shift of S4716. Also, the $(\delta\phi)_{prec}^{Sch}$ value $14'.8 \pm 1'.3$ proposed by \cite{2022ApJ...933...49P} falls within our estimated range of pericentre shift.

\section{Astrometric Capabilities and Detection of $f(R)$ Gravity Effects}\label{b2}
As we have seen in the previous section, the lighter and intermediate mass scalarons ($10^{-22}$ eV and $10^{-19}$ eV) are unscreened and the heavier scalarons ($10^{-16}$ eV) are screened near the GC black hole. The unscreened scalarons affect the amount of pericentre shift in different stellar orbits. In this section we investigate whether those pericentre shifts are detectable by the current and upcoming astrometric facilities.

\par The existing Keck telescope, the GRAVITY beam combiner at the Very Large Telescope (VLT) and the upcoming Thirty Meter Telescope (TMT) carry sufficient potential to test gravity near the GC black hole. The astrometric uncertainties of Keck, TMT and GRAVITY are respectively  $\sigma_{Keck}=0.16$ mas, $\sigma_{TMT}=0.015$ mas and $\sigma_{GRAVITY}=0.030$ mas\cite{2017PhRvL.118u1101H,2018A&A...615L..15G}. Following the work done by \cite{Zakharov2018}, we are considering an angle $\theta=2\sigma$ which is measurable with $95\%$ confidence level by these astrometric facilities. Therefore, we have calculated the amount of pericentre shifts of different stellar orbits (with a fixed high eccentricity 0.9). The results for unscreened scalarons ($10^{-22}$ eV and $10^{-19}$ eV) are shown in Table \ref{ta1}. We also show the pericentre shift of the orbit of S4716 \cite{2022ApJ...933...49P} arising from unscreened scalarons in Table \ref{ta2}. The astrometric size of the pericentre shift angle has been calculated by multiplying $(\delta \phi)_{prec}^{SC}$ and $(\delta \phi)_{prec}^{Sch}$ with the factor $\frac{a \sin (i)}{D}$ where, $i$ is the orbital inclination and $D$ is the distance to the GC black hole. For maximizing the result, $i=90^\circ$ and distance to the black hole in conformity with the present measurements has been taken as $8178$ pc \cite{refId0}.

\begin{table}[hptb]
	\caption{Astrometric shift of different stellar orbits in the sky arising from unscreened scalarons.}
    \centering
	{\begin{tabular}{@{}ccccc@{}} \hline \hline
			{\shortstack[m]{Scalaron Mass\\ ($M_{\psi}$)}} & {\shortstack[m]{Semi-major axis\\ ($a$)}} & 	{\shortstack[m]{Bounds on $\psi_0$\\ (Min-Max)}}&{\shortstack[m]{Astrometric shift \\for scalarons \\ $(\delta\phi)_{prec}^{SC} \times\dfrac{a}{D}$ ($\mu$as)}} & {\shortstack[m]{Astrometric shift\\for GR(Pure Schwarzschild case) \\$(\delta\phi)_{prec}^{Sch} \times\dfrac{a}{D}$ ($\mu$as)}} \\
			 
			\hline
			\multirow{3}{*}{} {$10^{-22}$ eV}  & \hphantom{0} $1000$ au & \hphantom{0}$0.01154 - 0.01877$ & \hphantom{0}$699.655 - 430.155$  & \hphantom{0} $497.386$ \\ \hline
			& \hphantom{0} $45$ au & \hphantom{0}$0.6378 - 1.0378$ & \hphantom{0}$697.895 - 428.905$ & \hphantom{0} $497.386$ \\
			$10^{-19}$ eV& \hphantom{0} $100$ au & \hphantom{0}$4.0253 - 6.5491$ & \hphantom{0}$699.692 - 430.055$ & \hphantom{0} $497.386$ \\
			& \hphantom{0} $1000$ au & \hphantom{0}$74.6127 - 121.39$ & \hphantom{0}$699.395 - 429.885$ & \hphantom{0} $497.386$ \\ 
			\hline
		\end{tabular} \label{ta1}}
\end{table}

\begin{table}[hptb]
	\caption{Astrometric shift of the orbit of S4716 in the sky arising from unscreened scalarons.}
	%\resizebox{\textwidth}{!}{ 
		{\begin{tabular}{@{}ccccc@{}} \hline \hline
				{\shortstack[m]{Scalaron Mass\\ ($M_{\psi}$)}} & {\shortstack[m]{Semi-major axis ($a$)\\ \& \\ Eccentricity ($e$)}} & 	{\shortstack[m]{Bounds on $\psi_0$\\ (Min-Max)}}&{\shortstack[m]{Astrometric shift\\ for scalarons \\ $(\delta\phi)_{prec}^{SC} \times\dfrac{a}{D}$ ($\mu$as)}} & {\shortstack[m]{Astrometric shift\\ for GR(Pure Schwarzschild case) \\$(\delta\phi)_{prec}^{Sch} \times\dfrac{a}{D}$ ($\mu$as)}} \\
				\hline
				$10^{-22}$ eV        & \multirow{2}{*}{\shortstack[m]{$a=407.29$ au \\ $e=0.74$}} & $0.0032 - 0.0052$ & $266.176 - 163.601$                          & \multicolumn{1}{c}{\multirow{2}{*}{$208.893$}} \\
				$10^{-19}$ eV        &                              & $84.66 - 137.74$  & $283.164 - 174.042$                          & \multicolumn{1}{c}{}    \\ \hline                 
				
			\end{tabular} \label{ta2}}
		%}
\end{table}

\par In Table \ref{ta1} and Table \ref{ta2} the range of $\psi_0$ and astrometric shift for scalarons (min-max) is realised from the bounds on $\gamma,\beta$. From the above tables, we observe that the values of pericentre shifts of the above stellar orbits due to the unscreened scalarons ($10^{-22}$ eV and $10^{-19}$ eV) are easily detectable by the present and upcoming astrometric facilities.

\section{Results and Discussions}\label{c}
From the variation of difference in pericentre shift between $f(R)$ gravity and GR (see Fig. \ref{fig1} \& Fig. \ref{fig2}) it is observed that for Schwarzschild and Kerr black holes the general relativistic pericentre shift starts dominating with increasing size of semi-major axis ($a$) for $R^2$ model and for eccentricity values, $e=0.1$ and $0.9$. The increase in GR pericentre shift is more prominent for $r_c =100$ au than $r_c=1000$ au. Therefore, increase in scale radius, $r_c$ flattens the rise of GR towards wide orbits. It is generally observed that $f(R)$ pericentre shift approaches the one in GR only at very small values of $a$. It means that the effect of $f(R)$ gravity is appreciable only in case of very compact orbits. Generality of this statement has been investigated in the model independent $f(R)$ scalaron gravity and the results are summarised below.

\par Table \ref{tab2} presents the allowed range of the scalaron field amplitude ($\psi_0$). It is seen that at $a=45$ au, $100$ au and $1000$ au the scalaron field amplitude falls in the range $\psi_0=10^{-6} - 10^{-2}$ for $M_\psi=10^{-22}$ eV. For the entire range of considered semi-major axis the scalaron field amplitude falls in the range $\psi_0=0.64-121$ for $M_\psi=10^{-19}$ eV. For $M_\psi=10^{-16}$ eV at $45$ au the scalaron field amplitude is of the order of $10^{-22}$. For $M_\psi=10^{-16}$ eV at $45$ au the scalaron field amplitude are extremely small and hence they represent scalarons with extremely large Yukawa coupling. Figure \ref{fig4} show the variation of $\psi_{0}/\phi$ with respect to PPN parameters. It is seen that the lighter scalarons ($10^{-22}$ eV) are screened for orbits with $a= 45$ au and $100$ au. These scalarons become unscreened at orbits of $a=1000$ au. The intermediate mass scalarons ($10^{-19}$ eV) are unscreened for the entire range of orbits ($45$ au – $1000$ au). But, the sufficiently heavier scalarons ($10^{-16}$ eV) are completely screened for all the orbits. In the orbit of the star S4716 ($a=407$ au, $e=0.7$), only the lighter scalarons ($10^{-22}$ eV) and intermediate mass scalarons ($10^{-19}$ eV) tend to remain unscreened. The scalaron field amplitude extracted from the observed pericentre shift of S4716 is found to be consistent with the estimated values of $\psi_0$ and pericentre shift displayed in Table~\ref{tab5} and obtained from the consideration of present bounds on PPN parameters. This suggests that measurement of any effect of the heavier scalarons ($10^{-16}$ eV) on the considered orbits is not possible. But, lighter ($10^{-22}$ eV) and intermediate mass ($10^{-19}$ eV) scalarons are unscreened and hence they may produce measurable effects. Although, the pericentre shift in $R^2$ gravity indicates its importance in high curvature region (compact orbit or in the very early universe as studied by Starobinsky \cite{1980PhLB...91...99S}), the model independent approach has shown that unscreening of modified gravity can be realised in low curvature regime too (wide orbits, $a \approx 1000$ au).

\par On further investigating the possibility of detection of $f(R)$ gravity effects, it is seen that for the unscreened scalarons ($10^{-22}$ eV and $10^{-19}$ eV) the GR (Schwarzschild) astrometric shift lies within the allowed range of astrometric shift for scalarons (see Table \ref{ta1}). Hence, the pericentre shift due to scalarons is comparable to GR pericentre shift below  orbital radii of $1000$ au. Also, it is evident that in the orbit of S4716 any effect due to these unscreened scalarons can be easily detected (see Table \ref{ta2}). The pericentre shift values for unscreened scalarons ($M_\psi=10^{-22}$ eV and $10^{-19}$ eV) are above the astrometric capabilities of the current Keck telescope ($\theta_{Keck} \approx 0.32$  mas), GRAVITY ($\theta_{GRAVITY} \approx 0.06$ mas) beam combiner at VLT and upcoming Thirty Metre Telescope (TMT) ($\theta_{TMT} \approx 0.03$ mas) and should be easily detectable through these facilities.

The inference drawn on screening of $f(R)$ gravity is based on present bounds on the PPN parameters $(\gamma,\beta)$. These are not as robust as the bounds realised in the solar system. Therefore, more stringent bound on these will enable one to further refine the study of screened modified gravity near the GC black hole. It is also realised that the future astrometric measurement of orbital pericentre shift are expected to possess sufficient potential for ruling out modified gravity theories.


%%% Comment out this section when you 
\bibliographystyle{mnras}
\bibliography{references}

\end{document}